\documentclass[pre,twocolumn,twoside,byrevtex,superscriptaddress,floatfix]{revtex4-1}

\usepackage{libs}
\usepackage{settings}
\usepackage[english]{babel}

\makeatletter
\adddialect\l@en\l@english
\makeatother

\setboolean{twocolswitch}{true}

\begin{document}

\title{\protect Ratioing the President: An exploration of public engagement with Obama and Trump on Twitter





}
\author{
\firstname{Joshua R.}
\surname{Minot}
}
\email{jminot@uvm.edu}

\affiliation{
Computational Story Lab, Burlington, Vermont, USA
}

\affiliation{
  Vermont Complex Systems Center, Burlington, Vermont, USA
}

\affiliation{
    Department of Mathematics and Statistics, University of Vermont, Burlington, Vermont, USA
}
\author{
\firstname{Michael V.}
\surname{Arnold}
}

\affiliation{
Computational Story Lab, Burlington, Vermont, USA
}

\affiliation{
  Vermont Complex Systems Center, Burlington, Vermont, USA
}

\affiliation{
    Department of Mathematics and Statistics, University of Vermont, Burlington, Vermont, USA
}


\author{
\firstname{Thayer}
\surname{Alshaabi}
}

\affiliation{
Computational Story Lab, Burlington, Vermont, USA
}

\affiliation{
  Vermont Complex Systems Center, Burlington, Vermont, USA
}

\affiliation{
    Department of Computer Science, University of Vermont, Burlington, Vermont, USA
}

\author{
\firstname{Christopher M. }
\surname{Danforth}
}
\affiliation{
Computational Story Lab, Burlington, Vermont, USA
}
\affiliation{
  Vermont Complex Systems Center, Burlington, Vermont, USA
  }
\affiliation{
    Department of Mathematics and Statistics, University of Vermont, Burlington, Vermont, USA
}

\author{
\firstname{Peter Sheridan}
\surname{Dodds}
}
\email{pdodds@uvm.edu}

\affiliation{
Computational Story Lab, Burlington, Vermont, USA
}

\affiliation{
  Vermont Complex Systems Center, Burlington, Vermont, USA 
  }
\affiliation{
    Department of Mathematics and Statistics, University of Vermont, Burlington, Vermont, USA
}

\date{\today}

\begin{abstract}
  \protect
  The past decade has witnessed a marked increase in the use of social media by politicians, most notably exemplified by the 45th President of the United States (POTUS), Donald Trump. 
On Twitter, POTUS messages consistently attract high levels of engagement as measured by likes, retweets, and replies. 
Here, we quantify the balance of these activities, also known as ``ratios'', and study their dynamics as a proxy for collective political engagement in response to presidential communications. 
We find that raw activity counts increase during the period leading up to the 2016 election, accompanied by a regime change in the ratio of retweets-to-replies connected to the transition between campaigning and governing. 
For the Trump account, we find words related to fake news and the Mueller inquiry are more common in tweets with a high number of replies relative to retweets. 
Finally, we find that Barack Obama consistently received a higher retweet-to-reply ratio than Donald Trump. 
These results suggest Trump's Twitter posts are more often controversial and subject to enduring engagement as a given news cycle unfolds. 
\\

\end{abstract}



\maketitle


\section{Introduction}
\label{sec:introduction}
The ability for US presidents to communicate directly with the public changed dramatically during the 20th and early 21st centuries. 
Moving from Franklin Delano Roosevelt's famous fireside chats through the television addresses of Harry Truman and John F. Kennedy, we now find ourselves in the era of social media campaigns and presidencies~\cite{frantzich_presidents_2018}. 
Communication technology has especially accelerated the ability for presidents to ``go public''~\cite{kernell_going_1997} and make appeals directly and instantly to the electorate.

With its introduction in the 2000s, social media provided a new platform for direct communication. 
New mechanisms for information sharing have consequences for contagion: rapid spreading of content through a user-base capable of responding in real-time. 
While these platforms democratize sharing for many categories of individuals on social media (e.g., celebrities interacting with fans),
US presidential accounts present a unique opportunity to analyze particularly salient signals indicative of broader sociotechnical phenomena in an increasingly prominent aspect of politics.

Advertisers and political campaigns alike are concerned with engagement metrics for social media posts. 
Twitter itself markets its data as an early warning system for customer satisfaction and reputation management~\cite{twitter_target_2019}---although the company has recently banned political advertising~\cite{ noauthor_political_nodate}.
The rate with which a given post garners interactions (e.g., clicks, profile views, mouse-hovers, etc.) is a common measure of engagement in the digital realm. 
Activities in response to social media posts include retweets/shares, likes/favorites, and replies/comments. 
These activities all reflect specific user actions such as endorsing a post or expressing a divergent opinion. 
Some have theorized that simple ratios of activities may be used as proxies for of the polarity of the public's response to a given message~\cite{isaac_ratio_2018, oneil_how_2017}.


The term `ratioed' is a Merriam-Webster ``word we're watching''~\cite{merriam-webster_words_nodate}.
On Twitter, the ratio value is generally taken to be defined by the ratio or balance of replies to likes or retweets. 
Here we will focus on the ratio of retweets to replies, as we show that like volume is often stable, while replies and retweets seem more reflective of specific public responses (Sec.~\ref{sec:results}).


In an effort to explore the dynamics of ratio space, in 2017 the website FiveThirtyEight Politics presented ternary ratios with activity counts normalized across retweets, comments, and likes~\cite{roeder_worst_2017}. 
Using this metric, the authors compared US politicians based on the ratio values their tweets receive. 
This work found noteworthy differences between politicians and political parties. 
As of late 2017, Trump tweets tended to be met with relatively more replies, while Obama tweets were met with relatively more retweets.
Tweets for both presidents had a high value of normalized activities that were likes. 
Beyond presidents, FiveThirtyEight Politics compared responses to the tweets of Republican and Democratic senators.
The senatorial accounts exhibited normalized response values largely reflective of their party's most recent president---Republican senators tended to have high values of normalized reply activities while Democratic senators had more retweets and likes.

\begin{figure*}[tp!]
    \centering
    \includegraphics[width=\textwidth]{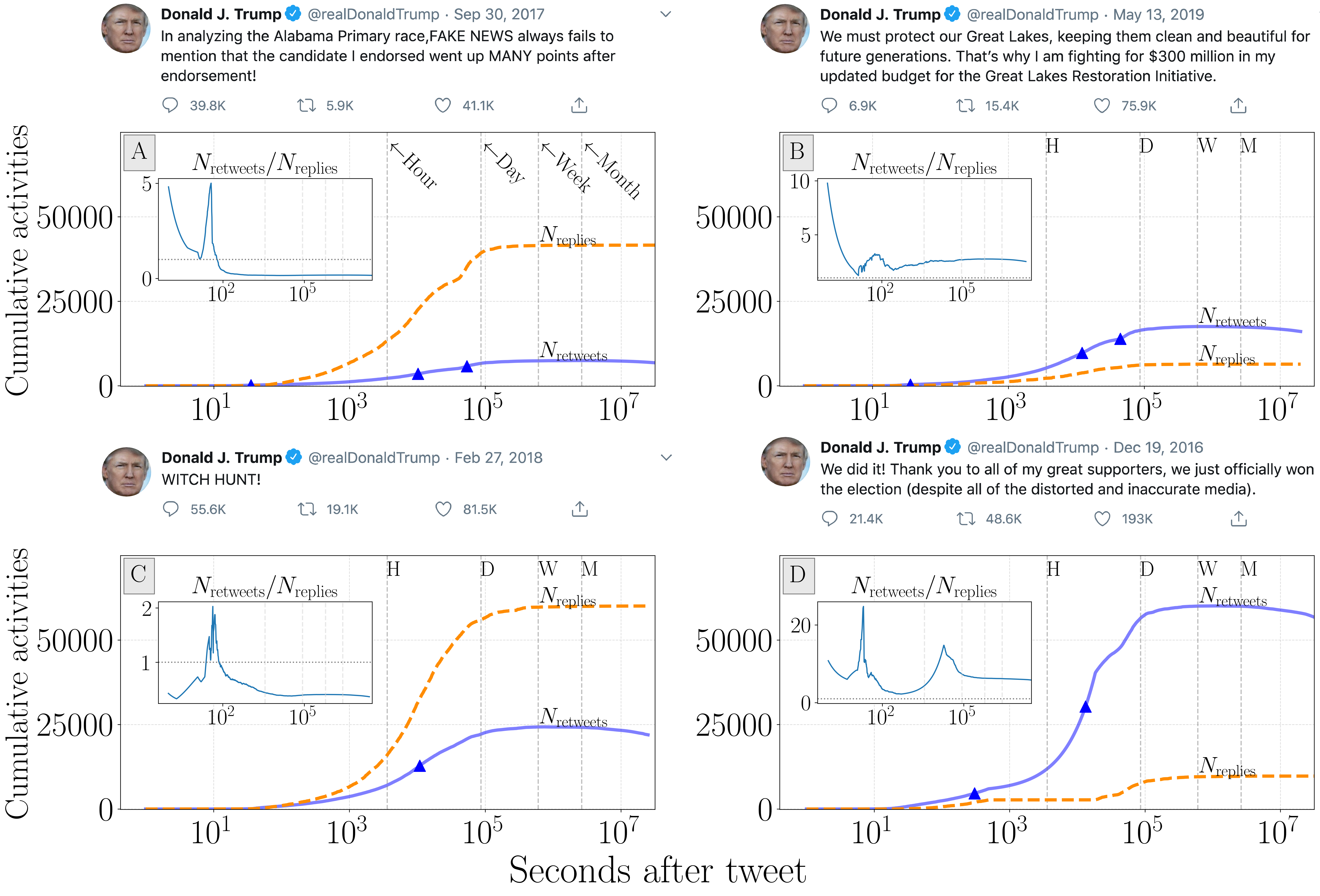}
    \caption{\textbf{Time series for the cumulative sum of retweet and reply activities} for four tweets authored by the Trump account after the 2016 presidential election.
    The temporal axis is logarithmically spaced to show early-stage growth. 
    Insets represent the cumulative ratio of $N_{\textnormal{retweets}}/N_{\textnormal{replies}}$ for the same time period. Vertical lines with \textit{H, D, W, M} correspond to hour, day, week, and month intervals.
    Triangles indicate inflection points where the second derivative of retweet volume transitions from positive to negative (see Sec.~\ref{sec:inflection_points} and Fig.~\ref{fig:knee_distributions}).
    The tweets examined here only provide some illustrative examples of how response activity time series behave. 
    Direct links to tweets for \href{https://twitter.com/realDonaldTrump/status/914269704440737792}{panel \textbf{(A)}: \url{https://twitter.com/realDonaldTrump/status/914269704440737792}}, \href{https://twitter.com/realDonaldTrump/status/1128051913419837442}{panel \textbf{(B)}: \url{https://twitter.com/realDonaldTrump/status/1128051913419837442}}, 
    \href{https://twitter.com/realDonaldTrump/status/968468176639004672}{panel \textbf{(C)}: 
    \url{https://twitter.com/realDonaldTrump/status/968468176639004672}}, and 
    \href{https://twitter.com/realDonaldTrump/status/810996052241293312}{panel \textbf{(D)}: 
    \url{https://twitter.com/realDonaldTrump/status/810996052241293312}}. 
    Tweet screenshots were collected on May 28, 2020.}
    \label{fig:tweet_ts}
\end{figure*}

More broadly speaking, time series of Twitter response activities have been used to investigate the relationship between on- and off-line political activity by analyzing correlations between current events and trends in cumulative activity sums~\cite{alashri_analysis_2016}. 
Fundamental models relating retweet activity to external activity have been proposed at the hour and day scale for the 2012 South Korean presidential election~\cite{ko_model_2014}. 
Kobayashi and Lambiotte find future retweet activity to be log-linear correlated with early tweet levels~\cite{kobayashi_tideh_2016}. 
Outside of Twitter, analysis of Instagram posts immediately preceding the 2016 US presidential election showed a higher volume of user activity for Trump supporters as well as more intense pro-Trump activity at the sub-day time-scale when compared to pro-Clinton posts~\cite{schmidbauer_2016_2018}.

Twitter specifically has been identified as playing an active role in the political arena as evidenced by studies of improved social organization during the Arab Spring~\cite{howard_opening_2011, tufekci_social_2012}, disinformation campaigns during the 2016 US presidential election~\cite{ruck_internet_2019, garrett_social_2019, guess_less_2019}, and political cohesion on the platform~\cite{cherepnalkoski_cohesion_2016, barbera_birds_2015}. 
Twitter has been shown to be a valuable source for data pertaining to domestic protests and social movements in the US~\cite{jackson_hijacking_2015, jackson_ferguson_2016, theocharis_using_2015}.
The 2016 US presidential election in particular has been extensively analyzed through the lens of Twitter data~\cite{bovet_influence_2019, lee_gendered_2016, darwish_trump_2017, quelch_twitter_2016}. 
There is also increasing academic interest in the unique communication style of President Trump on Twitter~\cite{clarke_stylistic_2019, ahmadian_explaining_2017}. 

Twitter been used to study several large-scale socio-technical phenomena such as the stock market~\cite{bollen_twitter_2011, bloomberg_bloomberg_2018} and medical research~\cite{hamed_twitter_2015, wetsman_how_2019}.
Messages on the platform have been shown to have a relationship with presidential approval rating polls~\cite{cody_public_2016, bovet_validation_2018, yaqub_analysis_2017, digrazia_more_2013, pasek_whos_2019, oconnor_tweets_2010}. 
Wang \textit{et al.} examine how likes on Twitter can identify the preferences of Trump-account followers~\cite{wang_catching_2016}.
Although noisy at times, Twitter data can provide valuable insights on how audiences and populations respond to current events and specific messaging. 


Information diffusion and cascades on Twitter and other social networks has been the topic of much research~\cite{hui_information_2012, banos_role_2013, bhattacharya_sharing_2012, goel_note_2015, crane_robust_2008}. 
Beyond politics on Twitter, Candia \textit{et al.} propose a bi-exponential model to describe collective attention for online videos and manuscripts~\cite{candia_universal_2019}. 
Media coverage of major events has been found to decay on roughly a weekly cycle, and scaling superlinearly with the population size of the affected area~\cite{prieto_curiel_temporal_2019}. 
Amati \textit{et al.} model retweet actions on dynamic and cumulative activity networks~\cite{amati_modelling_2016} while ten Thij \textit{et al.} identify retweet graph characteristics common among viral content~\cite{ten_thij_modelling_2014}.

Some researchers have proposed critical levels of activity required to trigger information cascades on Twitter in light of tweet characteristics~\cite{pramanik_modeling_2017}.
Others have proposed more general cascade requirements on social networks~\cite{kempe_maximizing_2003}.
Jin \textit{et al.} investigate how messages cross language and state communities on the platform~\cite{jin_can_2017}. 
Lee \textit{et al.} predict retweet volume based on prior retweet spacing, user meta-data, tweet content~\cite{lee_who_2015}. 
Others have identified celebrity-users (including Barack Obama) that are central in spreading new content~\cite{wu_who_2011}.


Regarding the representativeness of social media data, several studies have examined the degree to which Twitter represents the general population and voting public~\cite{barbera_understanding_2014, mellon_twitter_2017, wojcik_how_2019, murthy_urban_2016}.  
Grinberg \textit{et al.} investigate fake news exposure for Twitter accounts belonging to eligible voters and found disproportionate exposure to a small number of accounts~\cite{grinberg_fake_2019}.

How likely are Twitter users, and specifically POTUS account followers, to be eligible voters in the US? 
Pew Research Center has estimated that 26\% and 19\% of US adults follow the Obama and Trump accounts, respectively~\cite{wojcik_19_nodate}. 
The New York Times estimated however that less than 20\% of Trump's followers are voting-age US citizens~\cite{shear_how_2019}. 
These considerations are important to remember when attempting to establish a relationship between Twitter metrics and political outcomes. 
Nevertheless, the platform has been shown to be a powerful tool for evaluating public sentiment towards politicians, and even detecting adversarial actions in the US political process.


Presidential communications are a timely topic in light of rapidly changing norms surrounding how the executive branch communicates with the public. 
Moreover, highly influential Twitter users such as US presidents produce signals in the medium that largely transcend the social network dynamics. 
When a president tweets, the message is nearly instantly amplified and spread by official and unofficial sources on the platform. 
Indeed, US presidents are often discussed at rates approaching function words on Twitter~\cite{dodds_fame_2019}.
Examining ultrafamous users somewhat removes the more prominent effects of network topology on the response activity and provides an opportunity to view response behavior largely as a function of timing and message content.

The volume of response activities on Twitter is arguably the most tangible metric that is publicly available for evaluating user-base responses to content on the platform. 
Furthermore, these values are highly visible to users and may influence behavior when users seek to affect the collective response to a communication (e.g., `take a stand' by responding~\cite{jackson_hashtagactivism_2020}).
Establishing characteristic scales of activity volume and temporal dynamics for engagement is an important step in understanding how these values can provide insights on user-base response. 

Counts of user activities responding to tweets can be readily viewed through the icons at the bottom of every tweet, which also provide the interface for user engagement. 
These values provide the end user with an indication of the tweet popularity and, in some cases, controversiality.
Taken together with the cultural context surrounding the original tweet, the ratio of these values open up the possibility of studying tweets through the lens of ``ratiometrics''. 
This allows for the distillation of response activities into aggregated measures of the user base's reaction. 
Moreover, enables the comparison of response activities on Twitter across accounts and across time. 
From there, we can begin to use the ratio values as a criteria when evaluating the content of tweets. 

A simple calculation of the publicly accessible response activity counts is a starting point, but insufficient to fully investigate the full potential of ratiometrics.
Here we present a suite of tools---collectively referred to as the ``ratiometer''---that help us understand response activity profiles for tweets.
Many factors contribute to interpreting the activity counts for a tweet, including typical activity ratios for the user and the age of the tweet. 
Understanding the typical response that an account receives requires building a historical view of the user's tweets and their subsequent response activities. 
Another challenge is determining the typical response volume at a given time step since the tweet was issued.

We show example time series for retweets and replies in response to four Trump tweets authored after the 2016 election in Fig.~\ref{fig:tweet_ts}. 
In the case of an ultra-famous user like Trump, we see an immediate response to tweets with early retweets and replies occurring seconds after the original tweet. 
For Trump's tweets from after the 2016 election, we generally observe thousands of response activities within the quarter-hour, with response-activity counts nearing their final values within 24-hours of the original tweet. 
There may be modest growth during the proceeding week, but generally these values have stabilization periods at the day-scale~(Figs.~\ref{fig:snapshot_obama}~and~\ref{fig:snapshot_trump}). 
The insets in Fig.~\ref{fig:tweet_ts} show the retweet-to-reply ratio time series. 
These values often fluctuate even while individual ratio time series appear to have a more stable trend.   
The above highlights some of the challenges in building an understanding for the expected volume and time scale associated with response activities for a specific account. 

In the present study, we explore the time series of the volume of user activities in response to presidential tweets. 
We summarize the ratio values of responses to Obama and Trump tweets over three distinct periods of recent US political history. 
Our investigation includes ratio values for all three activity counts normalized against the sum of all activities (we refer to these as ternary ratios below).
We also present characteristic time scales for response activity volume over the three political periods for both presidents.
Finally, we present words that tend to appear more often in ratioed tweets for the $@$realDonaldTrump account. 
In section II, we describe our data and methodology. 
In section III, we present and discuss the results of our study. 
Finally, in section IV, we offer concluding remarks and point to areas for future work.  


\section{Methods}
\label{sec:methods}
\subsection{Data}

We construct the dataset analyzed in the present study by filtering Twitter's Decahose Streaming API \cite{twitter_decahose_2019} an, in principle, random 10\% sample of tweets authored since 2009. 
The stream contains a variety of activity types including tweets, retweets, quote tweets, and replies.
While the Twitter REST API serves up-to-date information for a given user or activity, the Streaming API data offers a snapshot of each activity at the moment of generation.
The specific form of this dataset allows us to create a historical timeline of activities; responses to each original activity arrive with timestamped counts for several metrics including replies, likes, and retweets.
For the purposes of this study, we define `activity' to refer to any user action recorded in the historical sample including original tweets, retweets, retweets with comments, and replies. 

Given the sampling rate of the Decahose API, each individual activity has an approximately 10\% chance of appearing in our data set. For popular users and tweets (i.e., tweets that garner retweets and replies that number in the thousands or more), we are almost certain to record activities responding to the original activity. Moreover, we are able to observe a number of activities with sufficient temporal resolution to construct historical time series down to the level of a single second.


\subsection{Tracking Retweets and Replies}

Starting with the 10\% random sample of Twitter activities, we collect activities responding to presidential tweets by filtering for replies and retweets responding to the $@$BarackObama and $@$realDonaldTrump accounts. 
We do not include the official US Presidential Twitter account ($@$POTUS) or the official White House account ($@$WhiteHouse). 
The random sample feed serves original tweets, retweets, and replies. 
Each of these activities are served in a tweet object that contains metadata which includes creation time and follower counts for the author of the activity. In the case of retweets, the tweet object contains a count of retweets and likes at the moment the activity is retweeted. 
Replies are tagged with user metadata and time of activity. 
Altogether, the metadata provides the historical data allowing construction of the response activity timeline. 


\subsection{Measuring the Ratio}
While retweets counts are observable via tweet metadata, replies must be gathered by maintaining a cumulative summation of reply activity counts (multiplying the final value by 10 to account for our sample rate). 
Validating these estimates for final reply values, we find the error to be less than 1\% in terms of actual/predicted reply counts for high activity tweets. 
In order to calculate the ratio at a common time step, we linearly interpolate the values for replies, retweets, and favorites, resulting in time series that can be sampled at arbitrary intervals for all activity types. 
We report time steps at the second resolution unless otherwise stated. 

There are several challenges associated with the historical feed, stemming from contradicting metadata on activity objects that occurred simultaneously.
With timestamps at the 1 second resolution, sometimes tweet activities are reported as occurring at the same time (inferring time stamps from Twitter snowflake IDs did not resolve these conflicts \cite{ twitter_twitter_2019}). 
Activities that reportedly occur at the same time often have conflicting values for retweets and likes (sometimes differing by hundreds of activities). 
This may be due the count values growing so quickly that the activities within a given second window have differing values and/or it may be an artifact of the slow update times within the platform's database. 
This latter point is made more notable when taken in combination with Twitter's practice of deleting millions of tweets per week (this would lead to fluctuations in the activity counts for a given tweet).

Regardless of the cause, fluctuations in activity counts result in time series that increase non-monotonically.
To make model fitting and data analysis easier, we remove observations that result in non-monotonically increasing behavior before the maximum value of the activity time series. 
When calculating first and second derivatives we use a Gaussian filter to smooth the time series so that we avoid incorrectly identifying noisy regions as notable transitions from positive to negative derivatives. 
There is decay in values for retweets while replies only increase---the exact reasons for the decay are not investigated here, but the lack of decay in the replies is due to the cumulative sum method we use for estimating reply values.
There is further work to be conducted on the possibility that these jagged portions are a signal of banned account activity, and the dynamics of the jagged regions may indicate activity by accounts that the platform ultimately deems ban-worthy.

Ternary activity values are calculated by dividing each activity count by the sum of all activities at a given time step. The ternary ratio value for activity type $\tau$ at time step $t$ is given by 

\begin{equation}
\mathcal{R}_\tau(t) = \frac{N_\tau(t)} { N_\textnormal{retweets}(t) + N_\textnormal{likes}(t) + N_\textnormal{replies}(t) } \,,
\end{equation}
where $N_\tau(t)$ is the count of the activity at time step $t$. With the above each observation is comprised of a 3-dimensional vector representing the normalized activity values for a tweet.

The ternary ratio values are represented on a ternary plot (2-dimensional simplex) where the values of activities at each time step sum to 1,

\begin{equation}
\sum_\tau \mathcal{R}_\tau(t) = 1\,.
\end{equation}

\subsection{Vocabulary of Ratioed Tweets}

We are also interested in how the text content of tweets relates to ratios of response activities. 
To investigate this we calculate rank turbulence divergence---as defined by Dodds \textit{et al.}---between ratioed and non-ratioed tweets \cite{dodds_allotaxonometry_2020}. 

The rank turbulence divergence between two sets, $\Omega_1$ and $\Omega_2$, is calculated as follows, 

\begin{equation}
\begin{aligned}
D^{R}_{\alpha}(\Omega_1 || \Omega_2) 
&= \sum \delta D^{R}_{\alpha,\tau}  \\
&= \frac{\alpha +1}{\alpha} \sum_\tau \left| \frac{1}{r_{\tau,1}^\alpha} - \frac{1}{r_{\tau,2}^\alpha} \right| ^{1/(\alpha+1)} \,,
\end{aligned}
\end{equation}
where $r_{\tau,s}$ is the rank of element $\tau$ ($n$-grams in our case) in system $s$ and $\alpha$ is a tunable parameter that affects the impact of starting and ending ranks.

Here, we take the content of tweets from the Trump account after his declaration of candidacy on June 16, 2015.
We then split this set by ratioed ($N_{\textnormal{retweets}}/N_{\textnormal{replies}} < 1$ ) and non-ratioed tweets ($N_\textnormal{retweets}/N_\textnormal{replies} > 1$). 
Obama's account did not have a sufficient number of ratioed tweets to conduct this analysis. 
From here we calculate the rank divergence, between the two sets.
We set $\alpha=1/3$ based on experimentation outlined in the original rank divergence piece by Dodds \textit{et al}. 
This value of alpha tends to balance the influence of high- and low-ranked items.

\section{Results and Discussion}
\label{sec:results}
\subsection{Overall time series}

\begin{figure*}[tp!]
    \centering
    \includegraphics[width=\textwidth]{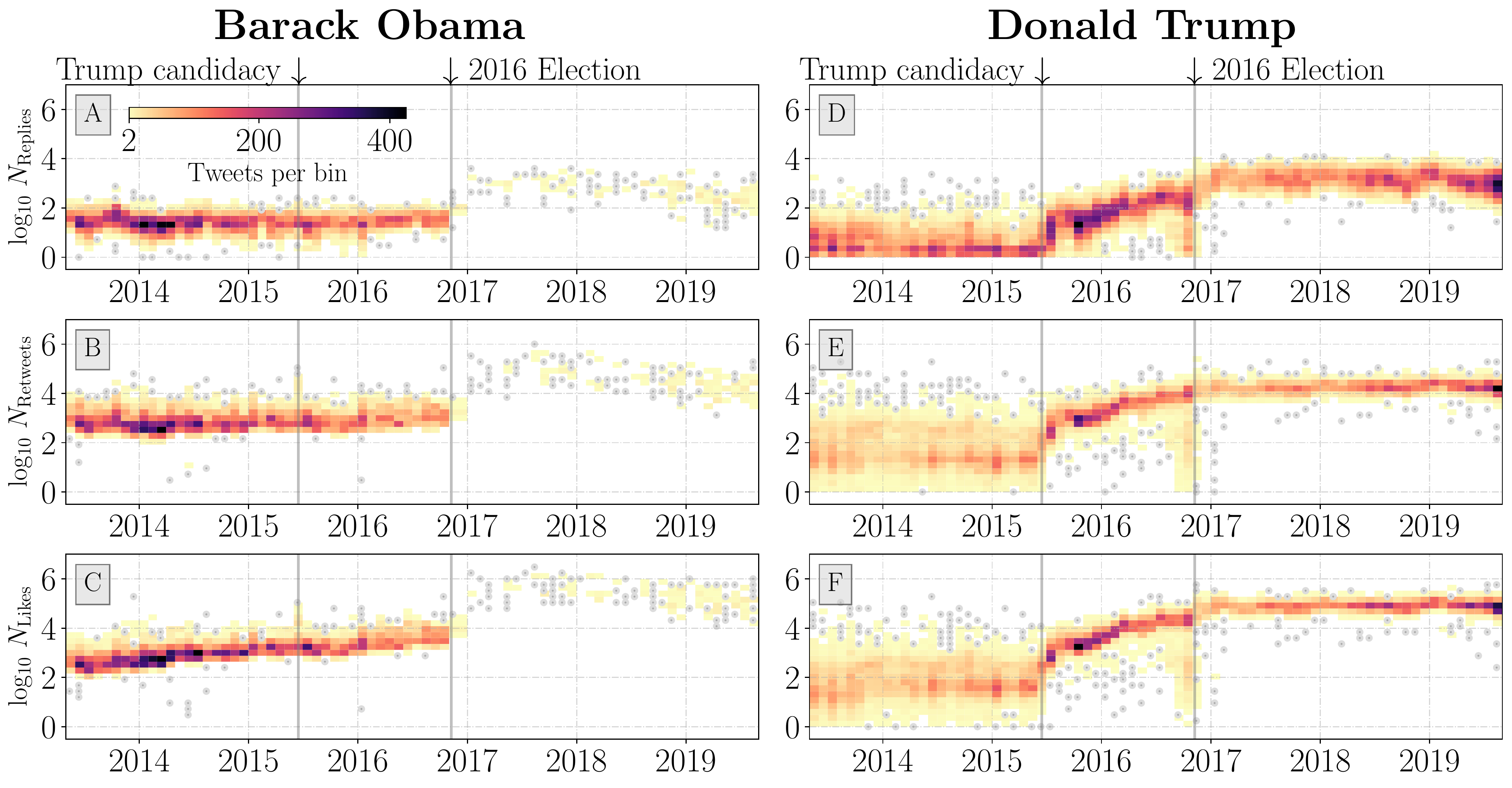}
    \caption{\textbf{Time series histogram of maximum activity counts for tweets} posted from the Barack Obama (\textbf{A}--\textbf{C}) and Donald Trump (\textbf{D}--\textbf{F}) Twitter accounts. 
    Each bin represents a collection of tweets at their original author date along with their respective maximum observed activity count.
    Bins with less than 2 tweets are shown as grey dots. 
    We include all tweet types (e.g., advertisements, promoted, etc.); see Section~\ref{sec:methods} for information on collection methods. 
    We annotate Trump's declaration of candidacy and the 2016 US general election with solid vertical grey bars. 
    A marked decrease in Obama account activity is apparent immediately following the 2016 election. 
    The region of outliers in the Trump time series immediately preceding the 2016 election has been determined to be largely reflective of promoted tweets which have abnormal circulation dynamics on the platform \cite{noauthor_what_nodate}.}
    \label{fig:all_time_series}
\end{figure*}

\begin{figure*}[tp!]
    \centering
    \includegraphics[width=\textwidth]{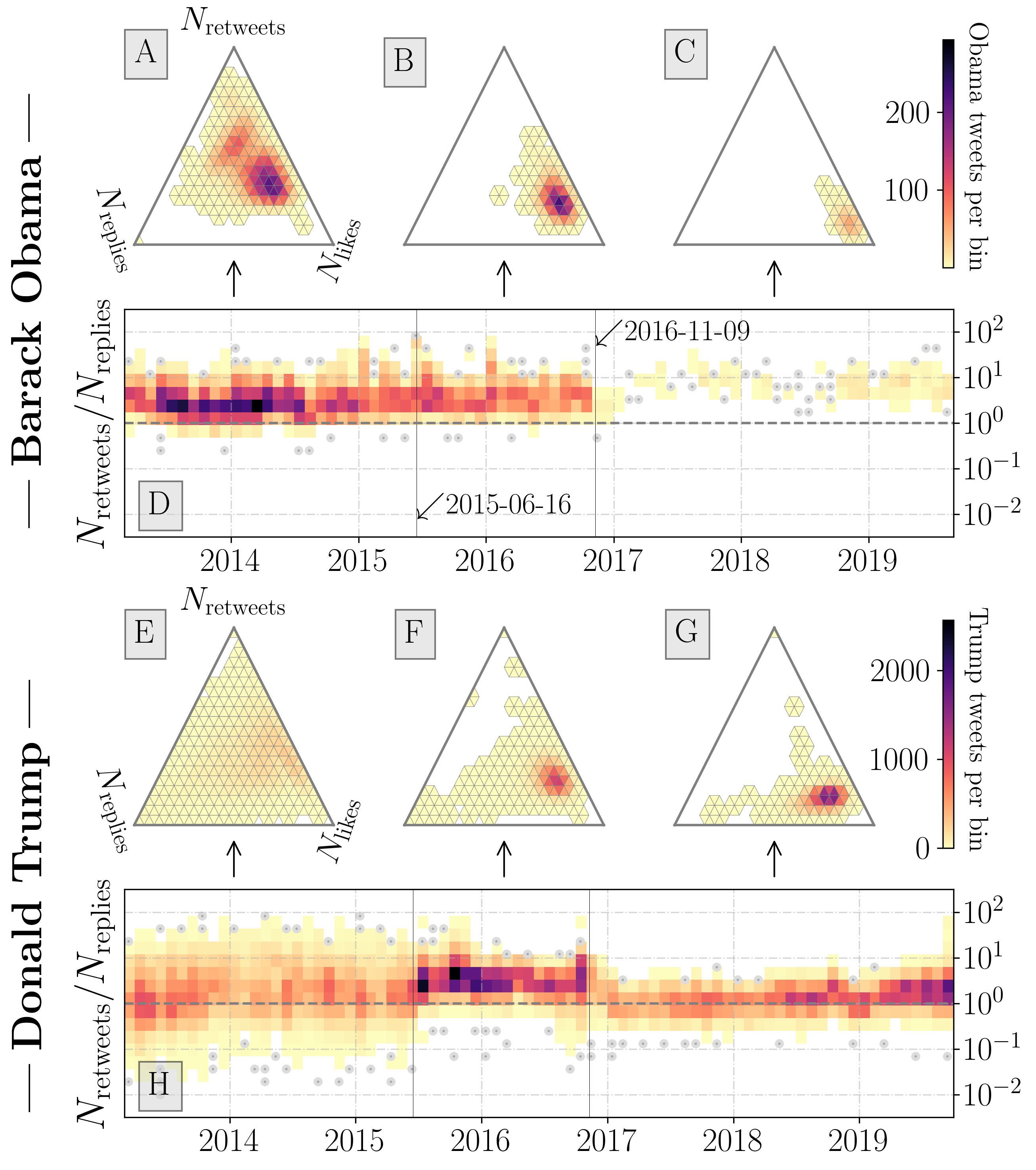}
    \caption{\textbf{Ternary histograms and $N_\textnormal{retweets}/N_\textnormal{replies}$ ratio time series} for the $@$BarackObama (\textbf{A--D}) and $@$realDonaldTrump (\textbf{E--H}) Twitter accounts. 
    The ternary histograms (\textbf{A--C} and \textbf{E--H}) represent the count of retweet, favorite, and reply activities normalized by the sum of all activities. 
    White regions indicate no observations over the given time period. 
    See Fig.~\ref{fig:trump_tern_timeseries} for examples of full time series for response activity for example tweets. 
    Heatmap time series (\textbf{D} and \textbf{H}) consist of monthly bins representing the density of tweets with a given ratio value. 
    Single observations (bin counts $<2$) are represented by grey points. 
    The two dates annotated correspond to the date of Trump's declaration of candidacy ($2015-05-16$) and the 2016 general election ($2016-11-09$). 
    We show the tendency for Trump tweets to have ternary ratio values with a greater reply component---with pre-candidacy tweets having higher variability and pre-election tweets having a higher $N_\textnormal{retweets}/N_\textnormal{replies}$ ratio value. 
    Post-election Obama tweets have ternary ratio values with more likes than other periods for both Obama and Trump.
    Screenshots were collected on May 28, 2020.}
    \label{fig:combined_tern_ts}
\end{figure*}{}

\begin{figure*}[tp!]
    \centering
    \includegraphics[width=\textwidth]{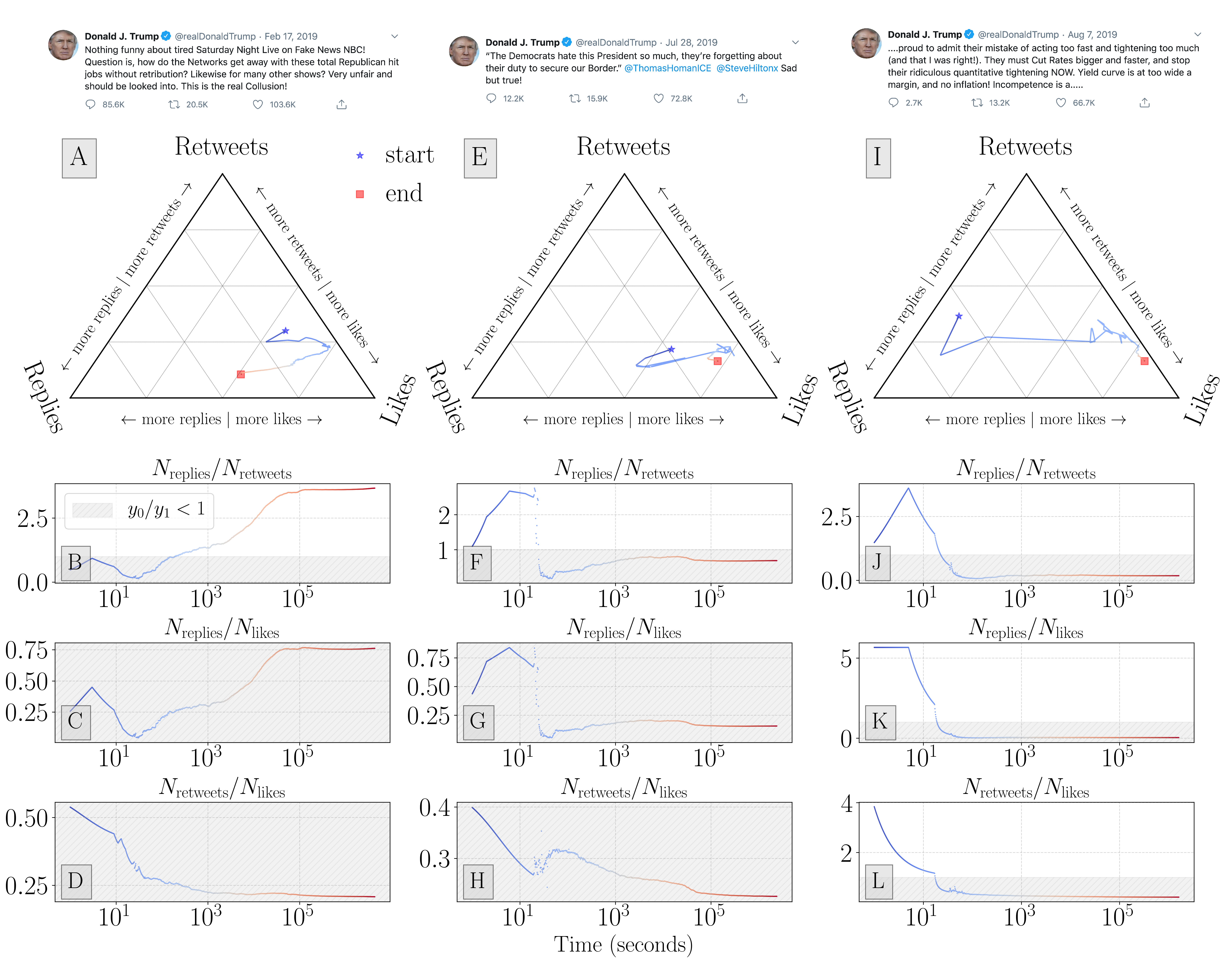}
    \caption{\textbf{Ternary time series for three popular Trump tweets}, selected to represent messages in approximately the upper 90th percentile, 50th percentile, and bottom 10th percentile of $N_\textnormal{retweets}/N_\textnormal{replies}$ ratios. 
    The time series represent observations from first activity observation to the final observation of the tweet. 
    These ternary time series contrast the simple 2-dimensional ratio trajectories and illustrate example trajectories through the 3-dimensional ratio space. 
    See Fig.~\ref{fig:snapshot_obama} and Fig.~\ref{fig:snapshot_trump} for the distribution of ternary ratio values for Obama and Trump tweets over time.
    Each of these three tweets were authored on May 25, 2019. 
    Direct links to tweets for \href{https://twitter.com/realDonaldTrump/status/1097116612279316480}{panels (\textbf{A--D})
    \url{https://twitter.com/realDonaldTrump/status/1097116612279316480}}, \href{https://twitter.com/realDonaldTrump/status/1155657137076527104}{panels (\textbf{E--H})
    \url{https://twitter.com/realDonaldTrump/status/1155657137076527104}}, and \href{https://twitter.com/realDonaldTrump/status/1159083364965654528}{panels (\textbf{I--L})
    \url{https://twitter.com/realDonaldTrump/status/1159083364965654528}}. 
    Screenshots were collected on May 28, 2020.}
    \label{fig:trump_tern_timeseries}
\end{figure*}

Viewing replies, retweets, and likes for individual tweets over time, we see how activities in response to Trump and Obama tweets have changed in the years around the 2016 election. 
In Fig.~\ref{fig:all_time_series}, we show  the final count (the activity volume 168 hours after the original tweet is authored) of activities for each tweet in the POTUS data set. 
Then-candidate Trump's tweets experienced a notable increase in response activity following his June 16, 2015 declaration of candidacy. 
Prior to this point, Trump's tweets garnered 2 to 3 orders of magnitude less activities than during the height of his campaign and his subsequent time in office. 

Because of the difficulty in sampling replies using our data (see Sec.~\ref{sec:methods} for tweets that receive few replies), many of the tweets from the Trump account from before his candidacy have a high degree of uncertainty in their true value. 
This uncertainty is introduced by our method for inferring reply counts introduces variance in the overall activity balance.


In Fig.~\ref{fig:combined_tern_ts} we present the normalized activity values for tweets from Obama and Trump from 2013 to 2019. 
The ternary histrograms allow us to compare all activity values, while the time series in Fig.~\ref{fig:combined_tern_ts}D and Fig.~\ref{fig:combined_tern_ts}H show the retweet-to-reply ratio. 
We choose the latter as it appears to offer a more descriptive measure of response activities (i.e., counts of likes are generally more stable within sub-regions of the yearly time series). 
The Obama $N_\textnormal{retweets}/N_\textnormal{replies}$ time series (Fig.~\ref{fig:combined_tern_ts}D) demonstrates how the Obama account tends to receive more retweets than replies (median ratio value of 3.67 between June 6, 2015 and November 8, 2016). 

Once Obama leaves office, the account receives consistently more retweets than replies on the limited number of tweets authored (median ratio value of 7.77 after November 8, 2016).

The Trump $N_\textnormal{retweets}/N_\textnormal{replies}$ time series (Fig.~\ref{fig:combined_tern_ts}H) shows the tendency of then-candidate Trump's account to be ratioed less during the campaign season. 
Soon after transitioning to office the Trump account begins receiving more replies relative to retweets---with a transition period roughly corresponding to the time between the day of the election and inauguration.   
For the campaign period Trump has a median ratio value of 3.1 while after the election the account garners a median ratio value of 1.32.

In Fig.~\ref{fig:combined_tern_ts}E we show a ternary histogram for the pre-campaign Trump account, where the high variance in reply estimates are evidenced by broad coverage of most regions of the simplex.
Error in estimating replies alone does not account for this high variance, with the like and retweet time series from Fig.~\ref{fig:all_time_series} showing higher variance for Trump as well. 
Another time series artifact is the presence of Trump tweets that were met with low activity counts around the 2016 election. 
Around the same time, the Obama account activity drops precipitously. 
After the election, Obama's limited number of tweets are met with consistently high counts of likes, retweets, and replies. 
The rapid increase in activities in response to Trump tweets, and the corresponding decrease in the overall variance of counts for activities, are important insights visible in Fig.~\ref{fig:all_time_series}. 
These measures are indicative of the meteoric rise of then candidate Trump, along with his now pre-eminent Twitter presence (his name now appears more frequently than the word ``god'' on most days ~\cite{dodds_fame_2019}).
While the two time periods are not directly comparable, by the end of the 2016 election, Trump's account consistently garnered more response activities than President Obama's. 
After the 2016 election, the Obama account's tweeting frequency is reduced while also experiencing a notable rise in response activities. 
These results helped inform the selection of distinct periods around the 2016 election. 
These time periods were both meaningful in a political context---marking Trump's declaration of candidacy and the 2016 election day---as well as in the context of response activity time series.

\subsection{Example Ternary Time Series}\label{sec:inflection_points}

\begin{figure}[tp!]
    \centering
    \includegraphics[width=\columnwidth]{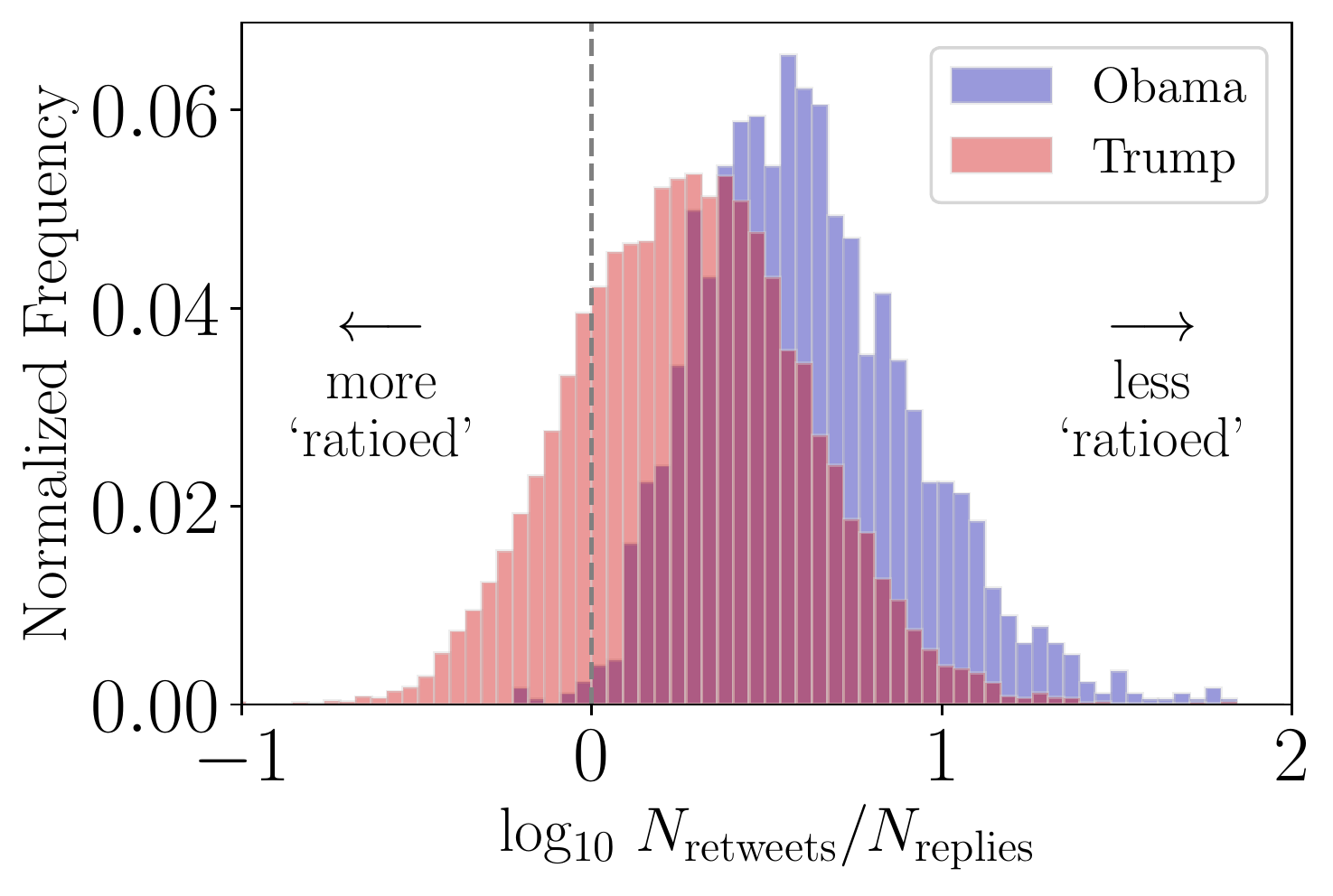}
    \caption{\textbf{Histogram of final observed ratios for tweets authored by $@$BarackObama and $@$realDonaldTrump accounts}. Observations left of the dashed vertical line correspond to tweets that are considered ratioed. 
    The shift in the distribution corresponding to Trump's account can be plainly seen---with the account producing more tweets that are ratioed than compared with Obama's account. 
    For both accounts, tweets shown here are restricted to those authored after Trump's declaration of candidacy. 
    Of 16,708 tweets included from the Trump account, 3,015 (18\%) have a $\text{log}_{10} N_\textnormal{retweets}/N_\textnormal{replies}$ value less than or equal to 0 and 13,693 (82\%) have a score greater than 0. Of 1,786 tweets from the Obama account, 7 ($<1\%$ have $\text{log}_{10} N_\textnormal{retweets}/N_\textnormal{replies}$ values $\leq 0$  while 1,779 ($>99\%$) have values $>0$.
    From the distribution of ratio values we see that ratioed tweets are outliers for the Obama account while the Trump account often gets ratioed and has a lower ratio value on average.}
    \label{fig:rep_rt_hist}
\end{figure}

\begin{figure*}[tp!]
    \centering
    \includegraphics[width=\textwidth]{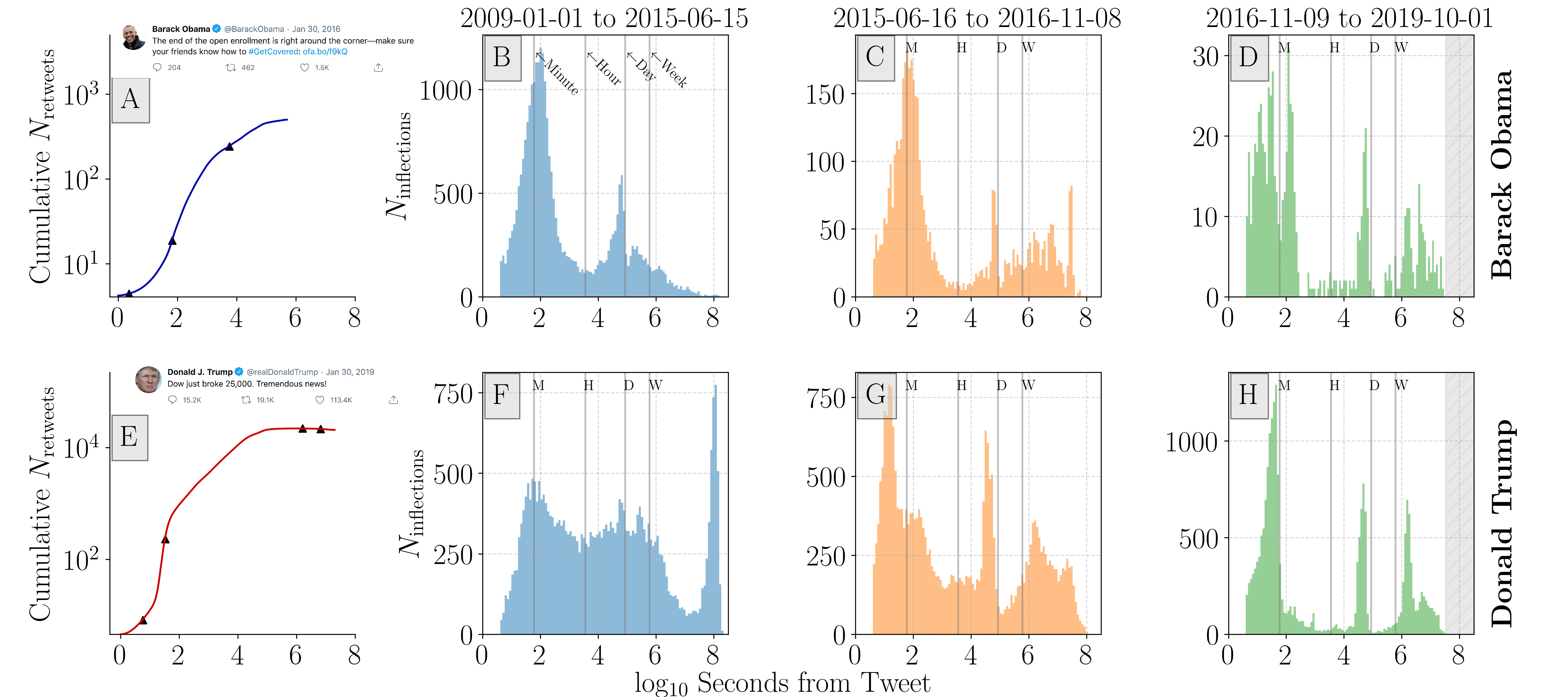}
    \caption{ \textbf{Distribution of inflection points}, where  
    $\frac{d^2}{dt^2} N_\textnormal{retweets} (t_{i-1}) > 0 \:  \textnormal{  and  } \:  \frac{d^2}{dt^2} N_\textnormal{retweet} (t_{i}) < 0 \,.$
    \textbf{A} and \textbf{E}: Example cumulative retweet time series and inflection points (solid triangles) for Obama and Trump tweets.
    Histograms show the distribution of inflection points across all tweets binned by time periods before (\textbf{B} and \textbf{F}), 
    during (\textbf{C} and \textbf{G}), 
    and after (\textbf{D} and \textbf{H}) 
    the 2016 US presidential election campaign. 
    The January 1st 2009 to June 15th 2015 period for Trump (\textbf{F}) contains inflection point counts that are largely reflective of low initial activity and high(er) late activity (months or years later) leading to unusually high values for seconds to first inflection point ($>10^8$ seconds). 
    For Obama's and Trump's time in office, tweets experience inflection points around 1-minute and 1-day after the tweet is authored---indicating characteristic time-scales of activity waning. 
    Direct links for 
    \href{https://twitter.com/BarackObama/status/693571153336496128}
    {Obama tweet (\textbf{A}): 
    \url{https://twitter.com/BarackObama/status/693571153336496128}} 
    and 
    \href{https://twitter.com/realDonaldTrump/status/1090729920760893441}
    {Trump tweet (\textbf{E}): 
    \url{https://twitter.com/realDonaldTrump/status/1090729920760893441}}.
    Screenshots were collected on May 28, 2020. 
    }
    \label{fig:knee_distributions}
\end{figure*}

\begin{figure}[tp!]
    \centering
    \includegraphics[width=\columnwidth]{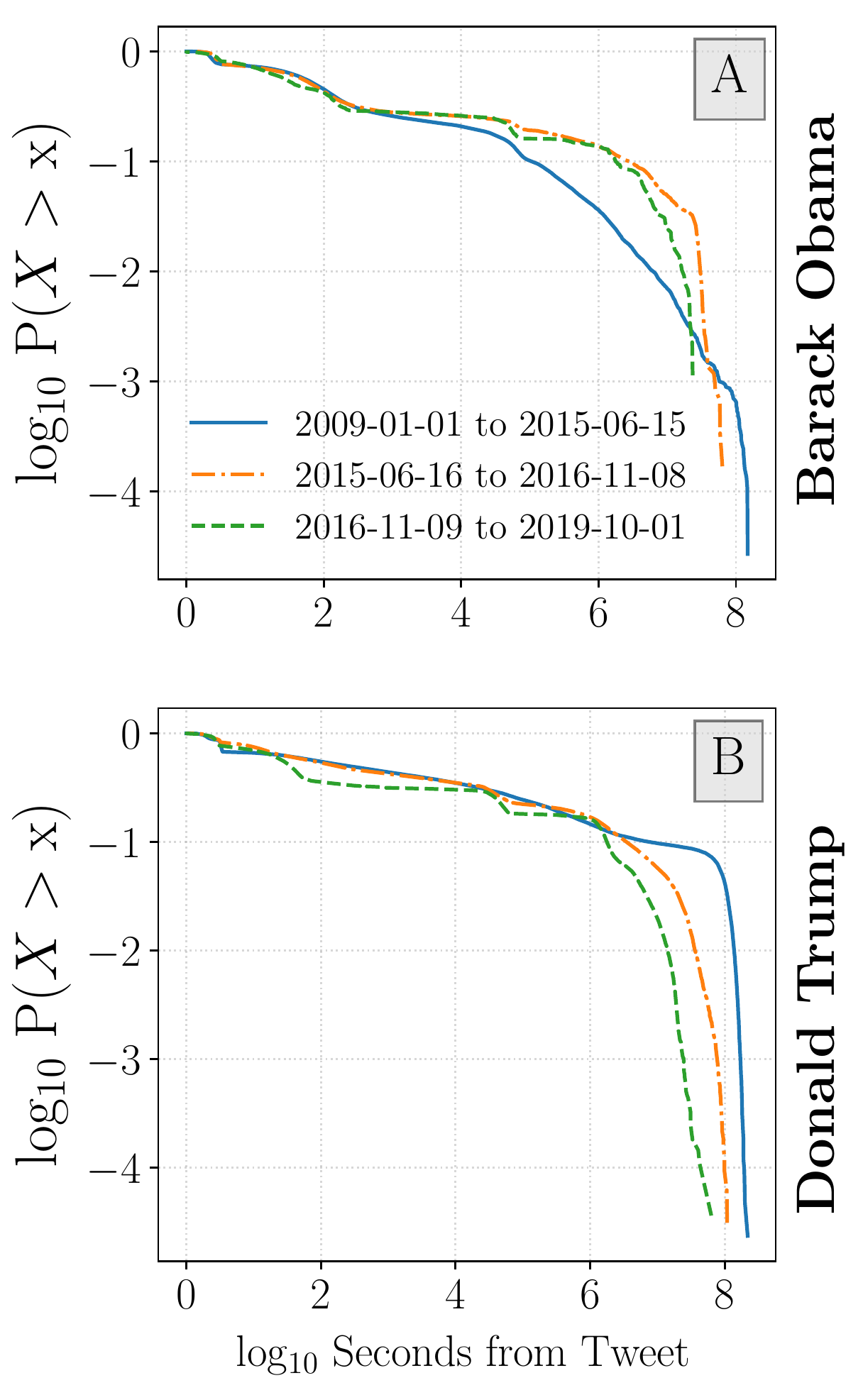}
    \caption{\textbf{Complementary cumulative distribution function for inflection point timing.} 
    Roughly 90\% of Obama inflection points (\textbf{A}) took place before $10^5$ seconds ($\sim 1$ day) for the period before the 2016 election cycle. 
    For the period during and after the 2016 campaign season, both Obama and Trump tweets (\textbf{B}) recieve 10\% of inflection points after roughly $10^6$ seconds ($\sim 10$ days)  from the initial tweet. }
    \label{fig:knee_ccdf}
\end{figure}{}

Using observations of responses to historical tweets, we can construct the time series for all ratios of activities. 
In Fig.~\ref{fig:trump_tern_timeseries} we show the ratio time series for three Trump tweets. 
We selected the tweets based on the value of their final retweet-to-reply ratio---with a tweet from approximately the 90th, 50th, and 10th percentiles of final $N_\textnormal{retweets}/N_\textnormal{replies}$ ratio values. 
The response activities to most tweets (with high response activity) tend to experience some early volatility, partially owing to the low number of observations.
One hour after the original tweet has been authored, the response ratios tend to fall into a stable region, or at least adopt a stable trend.
Within this signal there is also the effect of bots that are likely programmed to respond to Trump account activity within seconds. 

We show how ratios tend to stabilize by presenting ternary histograms of activity response ratios over the seconds and days after a tweet is published. Fig.~\ref{fig:snapshot_obama} shows how the Obama account largely has ratios biased towards retweets and likes throughout the period after a tweet is authored. 
Ratios for the Trump account (Fig.~\ref{fig:snapshot_trump}) tend to have greater variance in their ratio values in the seconds and hours after a tweet is released. 
The ratios for Trump's account are also biased towards more replies, relative to the Obama account, throughout the period after issuing a tweet. 
This is consistent with the final ratio values for each president across three political periods surrounding the 2016 election.

\subsection{Distribution of Ratios}

The final observed ratio value for a tweet offers an aggregate measure of user-base reactions. 
We examine tweets which were authored at least 168 hours prior to the activity calculation to ensure the responses have largely stabilized. 
Of course, there is still the possibility that users will respond to the tweet long after the original activity period, but we have found these lagging response activities to minimally affect the normalized ratio value. 
It is worth noting that this delayed response behavior is common for prominent users such as Obama and Trump, with some response activities taking place months or years after the original activity (often in reference to current events that are addressed in the old tweet: `there is always a tweet').

In Fig.~\ref{fig:rep_rt_hist} we show the distribution of Trump and Obama $N_\textnormal{retweets}/N_\textnormal{replies}$ ratios, highlighting the tendency for the Trump account to ``get ratioed'' more often relative to the Obama account. 
Of 13,639 tweets in Fig.~\ref{fig:rep_rt_hist} from the Trump account, $18\%$ receive more replies than retweets.
The Obama account generates a relatively limited number of tweets that receive more replies than retweets (less than $1\%$).
This simple ratio calculated with two of the three activity measures loses some context in terms of overall user-base responses, and we now move to incorporate the like activity volumes.

We subset the tweets for Presidents Obama and Trump based on the time periods introduced in Fig.~\ref{fig:all_time_series}. In Fig.~\ref{fig:combined_tern_ts} we show that for both Obama and Trump accounts there is a higher degree of variation in final ternary ratio values for earlier years. 
This is likely in part due to the reply thread mechanisms and structure changing on Twitter's platform, as well as the growing popularity of political accounts on Twitter, aside from background changes in how users engage on the platform. 
This shift is especially prominent for the Trump account, which experienced a marked growth in response activity after Trump's declaration of candidacy.   
The higher variance of final ratio values is accentuated when the sample of replies is lower, owing to our method of estimating reply activities. 
Comparing Obama and Trump ternary histograms (Fig.~\ref{fig:combined_tern_ts}) we observe more Trump activity ratios in the reply region of the simplex.

For the Obama account, Figs.~\ref{fig:combined_tern_ts}A--D, we see notably higher values for normalized retweet and like activity compared with the Trump account.
The retweet-to-reply ratio is consistently higher in the time series presented in Fig.~\ref{fig:combined_tern_ts}D. 
Once leaving office, the Obama account demonstrated a further tendency to receive a higher proportion of likes and retweets.



\subsection{Characteristic Time Scale}

To empirically investigate the characteristic time scale of response activities, we find points where the second derivative of retweet counts drops below 0,

\begin{equation}
\frac{d^2}{dt^2} N_\textnormal{retweets} (t_{i-1}) > 0 \:  \textnormal{  and  } \:  \frac{d^2}{dt^2} N_\textnormal{retweet} (t_{i}) < 0 \,.
\end{equation}
These points are viewed as being indicative of response activity starting to wane or `roll-over' as the rate at which new activities are generated decreases. 
Going forward we will refer to these points as inflection points (examples of these points can be seen with triangular markers in Fig.~\ref{fig:tweet_ts}). 

When viewing time since an original tweet in Fig.~\ref{fig:knee_distributions}, we see a higher density of inflection points around the 1 minute and 1 day time intervals. 
This suggests that many periods of intense activity take place within the first day of a tweet being authored. 
We also observe inflection points in cases where more sporadic engagement occurs later in the tweet response-activity timeline. 
This leads to another spike in activity around 1 week after the original tweet. 
Over 90\% of inflection points are observed before $10^5$ seconds ($\sim 1.15$ days) for both accounts and for all time periods as evidenced by complementary cumulative distribution functions (see Fig.~\ref{fig:knee_ccdf}).

There is a notable difference between the governing periods for Obama and Trump, Figs.~\ref{fig:knee_distributions}A and~\ref{fig:knee_distributions}G, respectively.
Trump tweet response time lines demonstrate a tendency to generate inflection points before the 1-minute mark---indicative of rapid response to his account's tweets.
Additionally, we find more time series inflection points after 1-week has passed for the Trump time series than compared with Obama. 
This suggests that users may return to Trump tweets later to comment on the content (perhaps as new political developments occur). 
This is further illustrated in Fig.~\ref{fig:knee_ccdf}B, where we show that 10\% of Trump inflection points take place after $10^6$ seconds ($\sim 10$ days). 
This is true for the pre-campaign, campaigning, and governing periods. 
By contrast, 90\% of Obama inflection points take place before $10^5$ seconds ($ \sim 1$ day). 

\subsection{Ratio Content}

\begin{figure*}[]
    \centering
    \includegraphics[width=\textwidth]{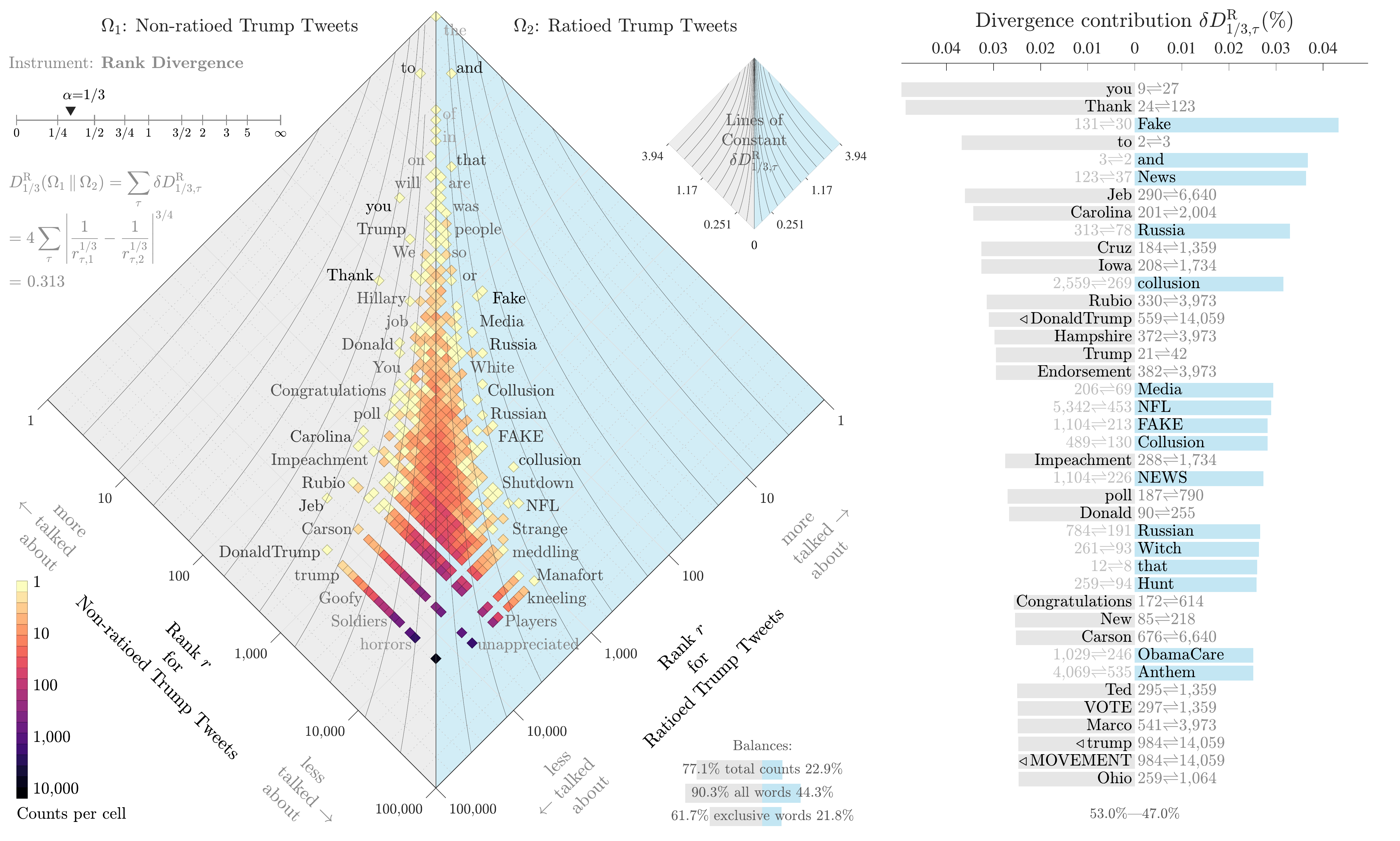}
    \caption{\textbf{Rank divergence allotaxonograph}~\cite{dodds_allotaxonometry_2020} for $1$-grams from Trump account tweets authored after Trump's declaration of his candidacy on June 16, 2015.
    ``Ratioed'' tweets are those where the proportion of retweets over replies is less than 1 ($N_\textnormal{retweets}/N_\textnormal{replies} < 1$), non-ratioed tweets accumulated a ratio greater than 1 ($N_\textnormal{retweets}/N_\textnormal{replies} > 1$). 
    There is a notable class imbalance. 
    The majority of tweets are non-ratioed, with a median value of $\sim 2$. 
    To generate the above figure we examined 3,313 ratioed tweets and 15,274 non-ratioed tweets. 
    The $1$-grams ``Fake'', ``News'', ``Russia'' and ``NFL'' are ranked higher in the ratioed corpus.
    For the non-ratioed corpus, the $1$-grams ``Jeb'', ``Carolina'', and ``Ted'' are ranked higher.
    This illustrates the tendency of ratioed tweets from this period to contain more politically contentious $1$-grams related to Trump scandals. 
    Non-ratioed tweets from this period more often contain campaign related messages. 
     }
    \label{fig:turb_trump}
\end{figure*}

We can investigate how tweet content relates to response activities by comparing the distributions of words that appear in ratioed and non-ratioed tweets. 
Taking the rank of the frequency of occurrence for each group, we then compare the rank-turbulence divergence for the two groups ~\cite{dodds_allotaxonometry_2020}. 
Per the allotaxonograph in Fig.~\ref{fig:turb_trump}, we are able to tune the impact of starting and ending rank magnitude on the overall score for the word.

In Fig.~\ref{fig:turb_trump}, we show that ratioed tweets contain more words related to fake news, the Mueller inquiry into Russian interference, and Colin Kaepernick and kneeling during the national anthem in the National Football League (NFL). 
For non-ratioed tweets, we see words from campaign-related communications (e.g., ``Jeb'', ``Iowa'', and ``VOTE'') tend to appear more often than in ratioed tweets. 
The imbalance of word counts in the ratioed and non-ratioed tweets is roughly proportional to the tweet imbalance with 22.9\% of words occurring in the ratioed tweets.
Of all unique words, nearly twice as many occur once or more in the non-ratioed tweets compared to the ratioed tweets.

\section{Concluding Remarks}
\label{sec:discussion}
We have considered how both the volume and kind of user activities in response to US presidents on Twitter varies over multiple time scales. 
We found the Trump account to have greater variability in the normalized ratio of activities---including a tendency to receive more replies relative to likes and retweets than when compared to the Obama account. 
Obama's tendency to receive more retweets and likes was only amplified after his departure from public office. 
We found that in ratioed tweets authored by Trump, words pertaining to fake news, the Mueller inquiry, and the NFL are more common than in non-ratioed tweets. 
We also showed more general results for response activity profiles, with responses to the two Twitter presidents often stabilizing around the 1-day scale. 
The Trump account also experiences more activity fluctuations after 1-week than compared to the Obama account, perhaps owing to users more often returning to Trump's comments as political actions unfold. 

How the public responds to messages over the course of a politician's career is a fundamental dynamic in politics. 
As candidates rise to prominence and are elected to office, the reach and importance of their communication changes. 
Further, the rapid proliferation of digital technology provides a backdrop of constant evolution in political communications. 
Response activity by users on social media provides a valuable set of features for gauging how social networks respond to political messaging. 
Beyond simple counts, response time series dynamics provide insight into the characteristics of political engagement. 
This includes determining when activity counts stabilize, detecting non-human user behavior, and perhaps establishing how polarized push-and-pull dynamics play out in response to messages. 
When combined with more conventional natural language processing techniques, these ``ratiometrics'' hold promise in improving our understanding of how specific messaging (e.g., word choice) affects user responses.


This study was not able to control for background changes in the Twitter interface---changes to the manner in which retweets are served could have effects on the response profiles. Further, there are some types of posts---namely ``promoted'' or ``ad'' posts---that have different distribution characteristics (i.e. may be viewed by a more targeted, limited audience). We were not able to fully filter these posts and they were included in the analysis. 

Twitter is a constantly changing system, and this is especially true for the nature of political discourse on the platform. 
While we make efforts here to highlight some notable exogenous events, we cannot control for background flux of user-engagement with politics. 
Simple changes that may be occurring include shifts in the distribution of users from across the political spectrum (i.e., more left- or right-leaning user activity) or more subtle changes in content (e.g., more spam and less in-depth conversation). 

With social media playing such a prominent roll in today's political process, it is our goal that the methods presented here can contribute to a broader suite of instruments for analyzing political communication in the digital realm. 
This is the first piece of an effort to systematically evaluate the communications of US presidents on Twitter. 
Building from our understanding of ratiometrics, the ``POTUSometer'' is envisioned as sitting alongside instruments such as the Hedonometer~\cite{dodds2011temporal} (which provides a measurement of collective sentiment on Twitter). 
The POTUSometer would tap into the `wisdom of the crowd' (or at least the reality of collective responses) in order to evaluate and better understand how presidents speak and are listened to on social media platforms.

Future research could explore how the sentiment changes with response activities (commented retweets and replies).
Similarly, topic modelling could be used to explore which subjects are discussed throughout the response activity time line. 
A more sophisticated null model for the ternary ratio values (and ratio time series) would be worthwhile to enable anomaly detection. 
There would be further work in the area of researching attention reinforcement along with analyses of time series stability. 
With recent advancements in language modelling, text regression on tweet content in order to predict ratio scores may be worthy of investigation. 
Finally, replicating this work across a broad base of users beyond Obama and Trump would better inform how certain social network attributes effect the above results. 


\acknowledgments 
We are grateful for funding from MassMutual and Google. We greatly appreciate the many helpful conversations with members of the Computational Story Lab. The support and guidance has been invaluable in exploring the topics we have outlined here. We are also grateful for the computational resources provided by the Vermont Advanced Computing Core.

\bibliography{revtex4}

\newpage



\newwrite\tempfile
\immediate\openout\tempfile=startsupp.txt
\immediate\write\tempfile{\thepage}
\immediate\closeout\tempfile

\setcounter{page}{1}
\renewcommand{\thepage}{S\arabic{page}}
\renewcommand{\thefigure}{S\arabic{figure}}
\renewcommand{\thetable}{S\arabic{table}}
\renewcommand{\floatpagefraction}{.9} 
\setcounter{figure}{0}
\setcounter{table}{0}

\section{Supplementary Material}

\begin{figure*}[ht!]
    \centering
    \includegraphics[width=.9\textwidth]{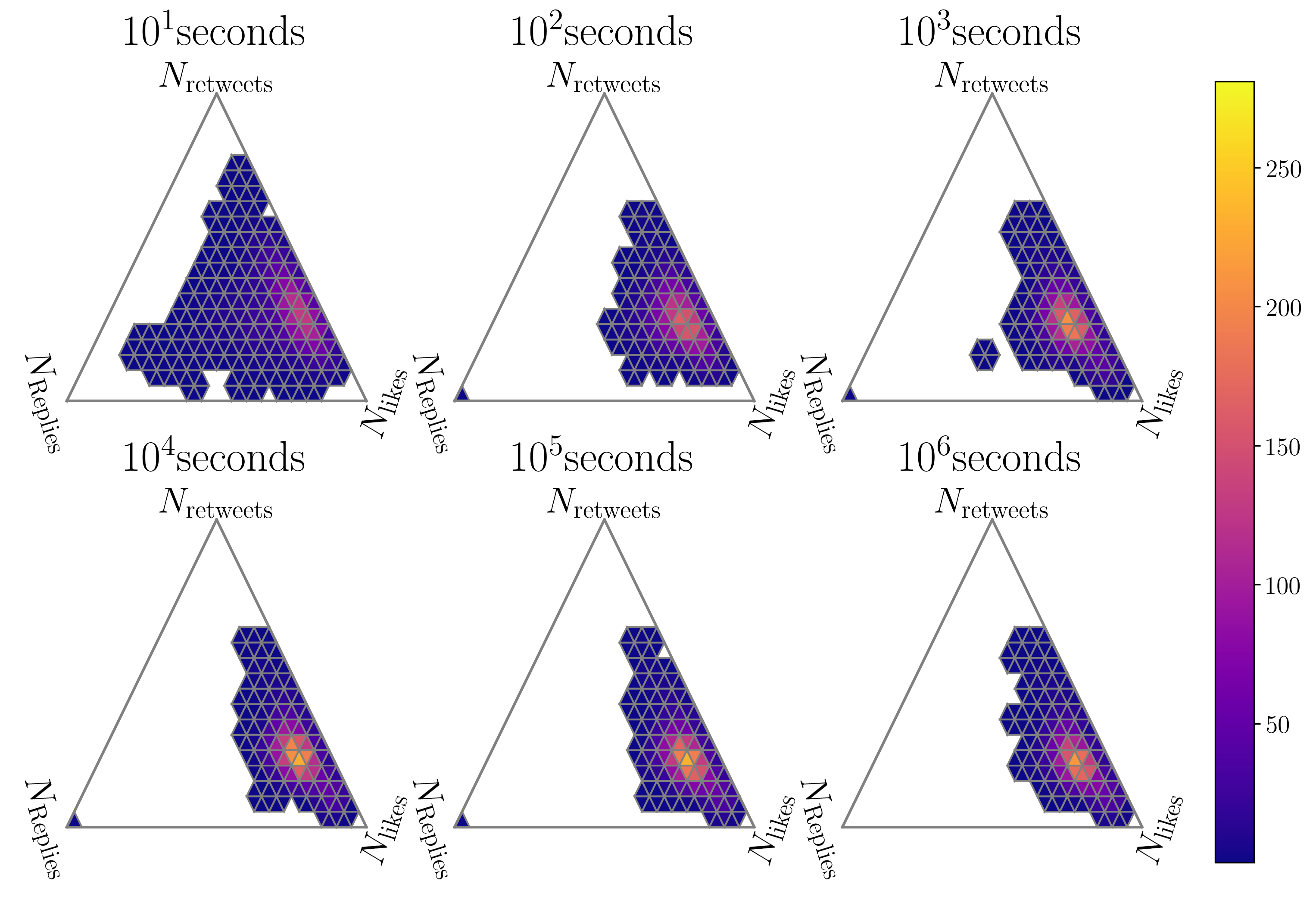}
    \caption{\textbf{Snapshots of ternary activity ratio values for tweets in response to the Obama account}. 
    Observations are recorded at logarithmically spaced intervals after the release of the original tweet. 
    Included here are 3,015 tweets from the period of time after Trump's declaration of candidacy on June 16, 2015. 
    Whereas Fig.~\ref{fig:combined_tern_ts} shows a final ratio value for tweets over distinct political periods, here we show how ratios unfold over the life of each tweet.  
    This is serves as a snapshot of the ternary ratio time series presented in Fig.~\ref{fig:trump_tern_timeseries}.
    Obama's tweets tend to receive a greater volume of likes and retweets relative to replies, and this ratio is often maintained throughout the response timeline.}
    \label{fig:snapshot_obama}
\end{figure*}{}

\clearpage

\begin{figure*}[ht!]
    \centering
    \includegraphics[width=.9\textwidth]{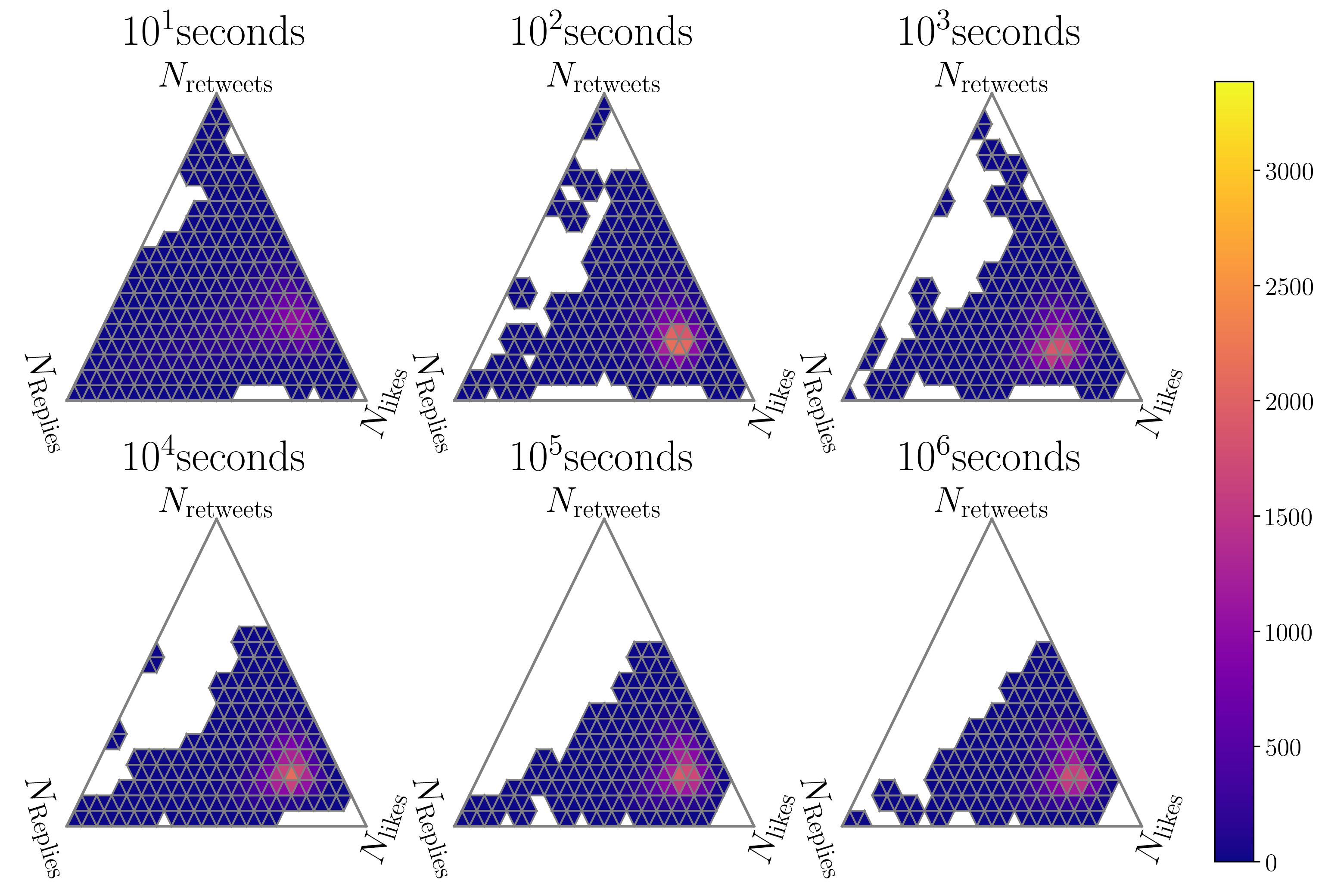}
    \caption{\textbf{Snapshots of ternary activity ratios for tweets in response to the Trump account}. 
    Observations are recorded at logarithmically spaced intervals after the release of the original tweet. 
    Included here are 16,708 tweets from the period of time after Trump's declaration of candidacy on June 16, 2015. 
    Whereas Fig.~\ref{fig:combined_tern_ts} shows a final ratio value for tweets over distinct political periods, here we show how ratios unfold over the life of each tweet.  
    This is serves as a snapshot of the ternary ratio time series presented in Fig.~\ref{fig:trump_tern_timeseries}.
    We can see the greater variation in ternary ratio values compared to the snapshots for Obama (Fig.~\ref{fig:snapshot_obama}).
    There is also a greater tendency for tweets to have higher reply counts when compared to the Obama snapshots.}
    \label{fig:snapshot_trump}
\end{figure*}{}

\label{sec:Appendix}

\end{document}